\documentclass[conference]{IEEEtran}
\makeatletter
\newcommand{\newlineauthors}{%
  \end{@IEEEauthorhalign}\hfill\mbox{}\par
  \mbox{}\hfill\begin{@IEEEauthorhalign}
}
\makeatother
\IEEEoverridecommandlockouts
\usepackage{hyperref}
\usepackage{cite}
\usepackage{pifont}
\usepackage{amsmath,amssymb,amsfonts}
\usepackage{algorithmic}
\usepackage{graphicx}
\usepackage{textcomp}
\usepackage{xcolor}
\usepackage{microtype}
\usepackage{tabularx}
\usepackage{array}
\usepackage{multirow}
\usepackage{makecell}
\usepackage{comment}
\usepackage{booktabs}
\usepackage{cleveref}
\usepackage{todonotes}
\usepackage{xspace}
\usepackage{colortbl}
\usepackage{siunitx}
\usepackage{accents}
\usepackage{tipa}

\newcolumntype{C}{>{\centering\arraybackslash}X}
\newcommand{\bwd}[1]{\accentset{\leftharpoonup}{#1}}

\newcommand{\zipactc}{ZIPA-CTC\xspace}

\newcommand{\powsm}{POWSM\xspace}

\newcommand{\pxeus}{PhoneticXeus\xspace}
\newcommand{\method}{SPAM\xspace}
\definecolor{colorint}{RGB}{214, 39, 40}
\definecolor{colorext}{RGB}{255, 176, 0}
\definecolor{colorother}{RGB}{44, 123, 182}
\crefname{section}{\S}{\S\S}
\Crefname{section}{\S}{\S\S}

\def\BibTeX{{\rm B\kern-.05em{\sc i\kern-.025em b}\kern-.08em
    T\kern-.1667em\lower.7ex\hbox{E}\kern-.125emX}}
\begin{document}

\title{Phone Segmentation and Recognition through Phonological Activation Mapping}

\author{\IEEEauthorblockN{Shikhar Bharadwaj\IEEEauthorrefmark{4}\IEEEauthorrefmark{1}\thanks{\IEEEauthorrefmark{1}These authors contributed equally. Order was determined randomly.},
Kwanghee Choi\IEEEauthorrefmark{2}\IEEEauthorrefmark{1},
Stephen McIntosh\IEEEauthorrefmark{3}\IEEEauthorrefmark{1},
Chin-Jou Li\IEEEauthorrefmark{4},
Eunjung Yeo\IEEEauthorrefmark{2},\\
Daisuke Saito\IEEEauthorrefmark{3},
Nobuaki Minematsu\IEEEauthorrefmark{3},
Shinji Watanabe\IEEEauthorrefmark{4},
Jian Zhu\IEEEauthorrefmark{5},
David Harwath\IEEEauthorrefmark{2} and
David R. Mortensen\IEEEauthorrefmark{4}}
\IEEEauthorblockA{\IEEEauthorrefmark{4}CMU, USA \,\, \IEEEauthorrefmark{2}UT Austin, USA \,\, \IEEEauthorrefmark{3}UTokyo, Japan \,\, \IEEEauthorrefmark{5}UBC, Canada\\
\{sbharad2,dmortens\}@andrew.cmu.edu \,\, \{kwanghee,harwath\}@utexas.edu \,\, \{smcintosh,mine\}@gavo.t.u-tokyo.ac.jp}
}

\maketitle

\begin{abstract}
Phone segmentation and recognition are inherently related tasks, yet modern approaches typically model them separately.
We argue that phonetic structure is already latent in the representations of self-supervised speech models (S3Ms), so one only needs to steer them to solve both tasks.
We leverage S3M-based Phonological Activation Mapping (SPAM), which maps each S3M representation frame to a vector of phonological feature activations, such as voicing and nasality.
On top of SPAM, we introduce two simple but effective lightweight, gradient-descent-free prediction heads: a recognition head and a segmentation head.
Our method requires less than a minute of phonetic transcriptions, and the recognition head generalizes to phones unseen during training.
Across a diverse range of datasets, our approach attains strong segmentation performance.
\end{abstract}


\section{Introduction}\label{sec:intro}
Phones are the smallest independent units of speech sound, conventionally transcribed across languages with the International Phonetic Alphabet (IPA).
Locating and recognizing them, \textit{i.e.}, phone segmentation and recognition, yields time-aligned phonetic transcriptions~\cite{garofolo1993timit,chodroff2024voxangeles}.
These are foundational to a range of applications, including clinical assessment of pathological speech, computer-assisted pronunciation training, and the documentation of endangered languages~\cite{shriberg2025clinical,Franco2010EduSpeakAS, prism,mortensen2021tusom2021, sun2026prosodic, mcintosch2026playground}.

However, obtaining such annotations is challenging, since expert phonetic transcription is slow and costly.
Annotating one hour of speech can take a trained phonetician roughly 40 to 100 hours~\cite{suchardt2025towards,seifart2018language}.
Transcriptions can also be subjective, with even experienced transcribers sometimes disagreeing on phone labels and segment boundaries~\cite{pitt2005buckeye,shriberg1991reliability}. 

These challenges have motivated automatic phone recognition and segmentation models.
Modern phone recognition is typically framed as a sequence-to-sequence task, much like automatic speech recognition (ASR)~\cite{xu22b_interspeech,zipa,powsm,bharadwaj2026pxeus}.
Accordingly, it is often trained with a connectionist temporal classification (CTC) loss~\cite{graves2006connectionist,xu22b_interspeech} or an attention-based encoder-decoder architecture~\cite{vaswani2017attention,chorowski2015attention,powsm}.
Phone segmentation, by contrast, is typically cast as frame-wise classification~\cite{strgar2023phoneme,zhu2022charsiu,xu2021explore,yeo2023speech}, labeling each frame as a boundary (1) or not (0).
Yet in human phonetic annotation, recognition and segmentation are performed together rather than as separate stages.
This raises a natural question: can recognition and segmentation be addressed through a unified representation?

We argue that they can, since the representations of self-supervised speech models (S3Ms)~\cite{baevski2020wav2vec,hsu2021hubert,chen2022wavlm} already encode sufficient information for both phone recognition and segmentation~\cite{choi2026b,choi2026self}.
In particular, S3M representation geometry is dominated by phonetic information~\cite{hsu2021hubert,choi2022opening, yossi2023phonemeanalysis,choi2025leveraging,choi24b_s3mphonetic}.

% https://docs.google.com/presentation/d/1HZf5IwXrh10riKUj5yn6fPeugmK91VyqR3H69PbaeDE/edit?usp=sharing
\begin{figure}[t]
    \centering
    \includegraphics[width=1\linewidth]{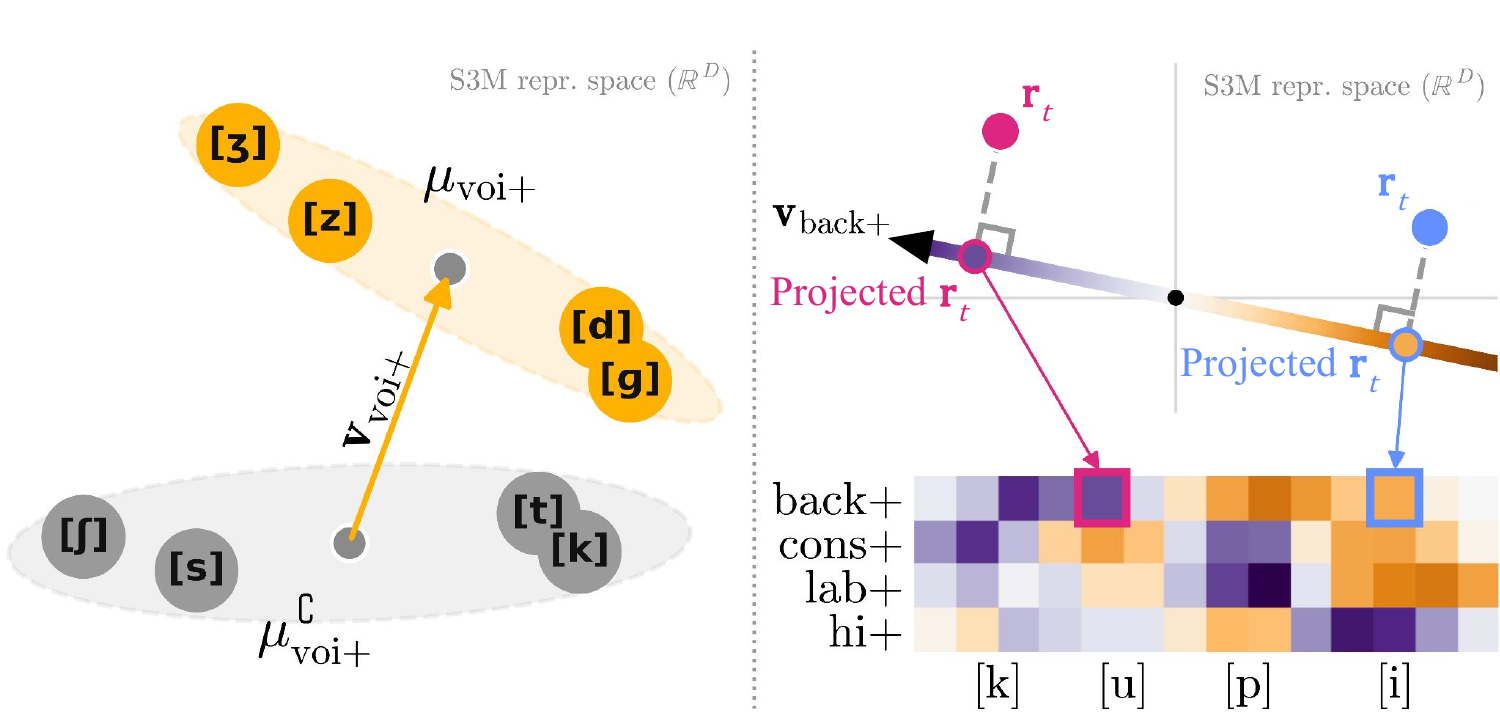}
    \caption{\textbf{Overview of S3M-based Phonological Activation Mapping (SPAM).} \textit{Left}: Phonological vectors~\cite{choi2026b, choi2026self} can be found in S3M representation space by taking the difference of means. For example, $\mathbf{v}_\texttt{voi+}$ is the difference between the mean representations of voiced phones and other phones.\\ \textit{Right}: 
    S3M frame representations $\mathbf{r}_t$ and $\mathbf{r}_{t'}$ from different timesteps $t$ and $t'$ are projected onto phonological vectors \cite{choi2026self,mcintosch2026playground} (\textit{e.g.}, $\mathbf{v}_\texttt{back+}$), yielding activation values (top right). Stacking these projections from different phonological vectors for each frame produces a time-aligned activation mapping (bottom right).
    }
    \label{fig:spam-overview}
    \vspace{-1em}
\end{figure}

To isolate phonetic information, we turn to \textbf{phonological feature decomposition}~\cite{mortensen2016panphon}.
Rather than treating each phone as an atomic label, phonological features describe phones through physical descriptions such as the presence of voicing, tongue height, and lip rounding.
In this framework, a phone becomes a vector of feature values. For example, [u] as in \textit{goose} is represented as \texttt{[voi+, hi+, back+, round+, \mbox{...}]}.
\texttt{PanPhon} is a library that maps over 5{,}000 IPA segments to 21 such features~\cite{mortensen2016panphon}.

Recent work~\cite{choi2026self,choi2026b} shows that these phonological features can be accurately modeled as linear directions, \textit{i.e.}, \textbf{phonological vectors}, in the representation space of some S3Ms.
For instance, adding the \textit{voicing vector} to a representation of [s] moves it toward [z] (\Cref{fig:spam-overview}, left).
Estimating these vectors is known to be highly sample-efficient~\cite{choi2026b}, a major practical advantage given the difficulties of expert annotation.
Projecting a frame's S3M representation onto these vectors yields an activation value for each phonological feature (\Cref{fig:spam-overview}, top right).
Stacking these phonological activations over time results in an \textbf{S3M-based Phonological Activation Map} \cite{choi2026self,mcintosch2026playground}, or \textbf{SPAM} (\Cref{fig:spam-overview}, bottom right).
This produces a time-aligned phonological representation of the speech.

We design a \textbf{segmentation head} that locates phone boundaries through dissimilarity between adjacent phonological activations, inspired by prior work on acoustic and raw S3M representations~\cite{glass1988multilevel,chang1997segmentation, kreuk2020selfsupervised}.
We also design a \textbf{recognition head} based on SPAM, which simply predicts phone identity at a frame from its phonological activations. 
Accordingly, we can predict any cataloged phone in \texttt{PanPhon}, even if it does not exist in the training data.
% We use \texttt{PanPhon}, which catalogs over 5000 IPA segments, for mapping speech to phones.

We claim that SPAMs have some key advantages for phonetic tasks like segmentation and recognition.
They are based on a naturally occurring representation subspace aligned with human linguistic abstractions, which generalizes across many languages. 
Therefore, SPAM provides interpretability, generalization, and sample efficiency by design.

\begin{figure}[t]
\centering
  \includegraphics[width=0.99\linewidth]{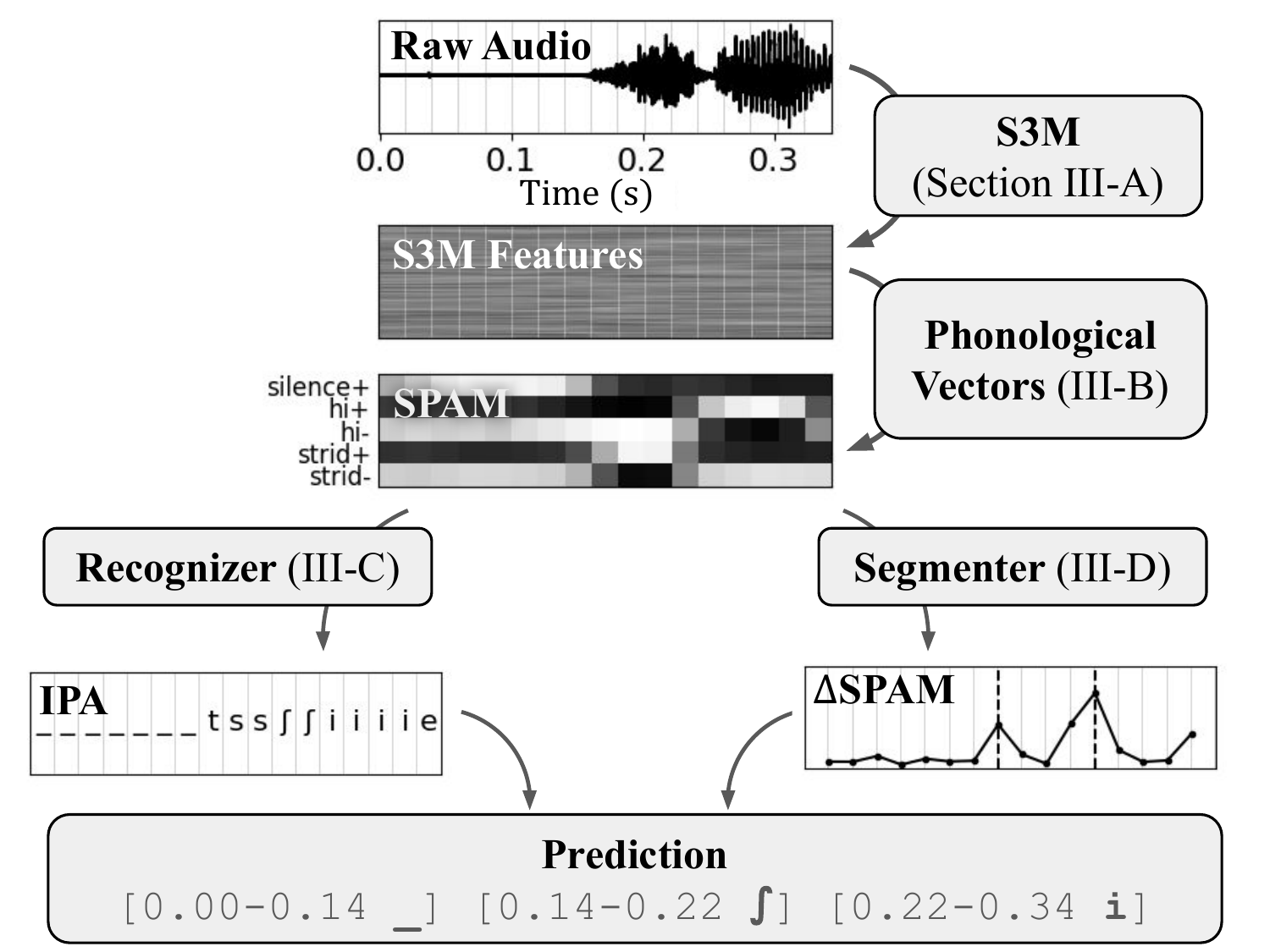}
  \vspace{-1em}
  \caption{
    Overview of our method (\Cref{sec:method}).
  }
  \label{fig:fig1}
  \vspace{-1em}
\end{figure}

Our contributions are as follows:
\begin{itemize}
\item \textbf{Novel segmentation and recognition methods.} We design two interpretable and lightweight prediction heads that perform both tasks directly on SPAM (\Cref{sec:method}, \Cref{fig:fig1}).
\item \textbf{Generalization and sample-efficiency.} Our method needs less than a minute of labeled data (\Cref{ss:abl-eff}), recognizes phones never seen during training (\Cref{ss:abl-comp}), and generalizes well to low-resource languages and atypical speech (\Cref{sec:expr}).
\item \textbf{Comprehensive evaluation.} We benchmark across a wide array of datasets and languages, reflecting realistic downstream use cases (\Cref{sec:expr}).
\end{itemize}

\section{Related work}
\noindent \textbf{Segmentation.}
Classic methods place boundaries at peaks of frame-to-frame acoustic dissimilarity~\cite{glass1988multilevel}, with mel-spectrogram dissimilarity still a surprisingly strong baseline
\cite{yang2024simple}.
Recent works use learned representations such as S3M~\cite{strgar2023phoneme} or self-supervised
CPC~\cite{kreuk2020selfsupervised,bhati2022unsupervised}, and HMMs over S3M features~\cite{yang2024simple}.
However, all stop at boundary detection, whereas we also explore phone recognition.

\noindent \textbf{Recognition.}
Modern multilingual phonetic recognizers, such as Allosaurus~\cite{li2020universal}, Allophant~\cite{glocker23_interspeech}, Wav2Vec2Phoneme~\cite{xu22b_interspeech}, ZIPA~\cite{zipa}, and PhoneticXeus~\cite{bharadwaj2026pxeus}, often learn a CTC, transducer, or encoder--decoder from large phone-transcribed datasets or grapheme-to-phoneme (G2P) pseudo-labels.
Allophant is the closest to our recognition head, using articulatory attributes.
Ours, in contrast, is gradient-descent-free, choosing each phone by matching its \texttt{PanPhon} feature vector against SPAM, requiring much less data.

\noindent \textbf{Segmentation and recognition.}
Chang and Glass~\cite{chang1997segmentation} propose ``segmentation by recognition,'' producing phone boundaries through segment-based recognition rather than purely local features.
More recent text-independent phone-to-audio alignment systems include Wav2Vec2-FC, Wav2Vec2-FS~\cite{zhu2022charsiu}, and TIPAA-SSL~\cite{tits2024textindependent}.
Wav2Vec2-FC uses frame-level cross-entropy (FCE), Wav2Vec2-FS uses forward-sum alignment without frame-level labels (similar to CTC), and TIPAA-SSL combines CTC with FCE.
In contrast, we use no training loss, relying on the phonetic structure SPAM already exposes.

\section{Method}
\label{sec:method}
% Harwath: We can also consider HMM-style method, such that it can do both recognition and segmentation together. We can consider phonotactic constraints, or only consider transition/self-loop probabilities.
 
\subsection{S3M-based Phonological Vectors}
\label{sec:prelim}

\noindent \textbf{S3M representation.}
A self-supervised speech model (S3M) encodes a waveform into a sequence of $D$-dimensional frame-level representations $\mathbf{R} = (\mathbf{r}_1, \dots, \mathbf{r}_T)$, where $\mathbf{r}_t \in \mathbb{R}^{D}$ and $T$ is the number of frames.
These can be read from any transformer layer, although this choice matters since phonetic information is distributed unevenly across layers~\cite{pasad2021layer}.
Following~\cite{choi2026self,choi2026b}, we primarily use the final layer of WavLM-large~\cite{chen2022wavlm} for our experiments. Model and layer ablations can be found in Appendix \ref{ss:abl-s3m}.

\noindent \textbf{Pooling.}
We extract a single vector for each phone, which typically spans several temporal frames,
using center pooling: the phone is represented by the frame at the center rather than by averaging over the span~\cite{choi24b_s3mphonetic,pasad2025speech}.
Center-pooled representations are preferable for extracting phonological vectors, as phonetic arithmetic~\cite{choi2026b} holds more reliably on them than on average-pooled representations~\cite{choi2025leveraging,choi2026self}.

% https://docs.google.com/presentation/d/1AzlxCgYgG9mIx45wLcHpniDqhZqvsWKtaRyEuFb7i10/edit?usp=sharing

\noindent \textbf{Phonological vectors.}
Recent work~\cite{choi2026b,choi2026self} shows that, within an S3M's representation space, each phonological feature can be effectively modeled as a linear direction recoverable as a difference of means.
We use \texttt{PanPhon}~\cite{mortensen2016panphon} to assign each phone its phonological features.
Note that \texttt{PanPhon} uses a ternary feature system with $+/0/-$.
$+$ and $-$ mark a property's presence or absence (\textit{e.g.}, voiced versus voiceless), and $0$ marks that the property does not apply to the phone (\textit{e.g.}, tenseness, which describes only vowels, is $0$ for consonants).

We split each ternary feature into two binary channels: \texttt{feature+} and \texttt{feature-}~\cite{choi2026b}, resulting in a set of phonological channels $C$. These are $1$ when the feature is $+$ or $-$, respectively, and $0$ otherwise.
Pooling all training utterances, let $\mathcal{S}_i$ denote the set of all phone segments where channel $i \in C$ is $1$ (\textit{i.e.}, active). We denote its complement by $\mathcal{S}_i^\complement$. For example, $\mathcal{S}_\texttt{voi+}$ contains all voiced segments, and $\mathcal{S}_\texttt{voi+}^\complement$ contains all other segments.
We discard channels whose positive or complement set is empty for the training vocabulary.

Let $c(s)$ represent the center frame of segment $s$.
The phonological vector for channel $i$ is a difference of means:
\begin{align}
  \mathbf{v}_i = \mathbb{E}_{s \in \mathcal{S}_i}\left[\mathbf{r}_{c(s)}\right] - \mathbb{E}_{s \in \mathcal{S}_i^\complement}\left[\mathbf{r}_{c(s)}\right]= \boldsymbol{\mu}_i - \boldsymbol{\mu}_i^\complement.
  \label{eq:phonovec}
\end{align}
For example, we estimate two phonological vectors for the voicing feature: $\mathbf{v}_\texttt{voi+} = \boldsymbol \mu_\texttt{voi+} - \boldsymbol \mu_\texttt{voi+}^\complement$ and $\mathbf{v}_\texttt{voi-} = \boldsymbol \mu_\texttt{voi-} - \boldsymbol \mu_\texttt{voi-}^\complement$.
% Note that segments, not phones, are equally weighted.

\subsection{S3M-based Phonological Activation Mapping (SPAM)} \label{sec:phone-posterior}
Using these phonological vectors, we turn the frame-wise S3M representations $\mathbf{r}_t$ into an S3M-based Phonological Activation Mapping (SPAM).
Recent works \cite{choi2026self,mcintosch2026playground} use this for visualization, whereas we formally define it and apply it to phone segmentation and recognition.

\method measures the similarity between each phonological vector and each representation, yielding a phonological activation vector $\mathbf m_t \in \mathbb{R}^{|C|}$.
Specifically, we project each frame-wise representation $\mathbf{r}_t$ onto each phonological vector $\mathbf{v}_i$ and apply a per-channel affine normalization, so that the different channels share a common range of values:
\begin{align}
  m_{t,i} &= \frac{\gamma}{\lambda_i} (\mathbf{r}_t^{\top}\mathbf{v}_i - \alpha_i),
  \label{eq:posterior}
\end{align}
where $\gamma$ is a scaling constant\footnote{We fix the scaling constant $\gamma=4$ for all of our experiments.} and $\alpha_i, \lambda_i$ are, respectively, the midpoint between and difference of the projected means:
\begin{align}
    \alpha_i = \tfrac{1}{2}\left(\boldsymbol{\mu}_i^{\top}\mathbf{v}_i
+ \boldsymbol{\mu}_i^{\complement\top}\mathbf{v}_i\right), \qquad
    \lambda_i = \boldsymbol{\mu}_i^{\top}\mathbf{v}_i- \boldsymbol{\mu}_i^{\complement\top}\mathbf{v}_i.
\end{align}
Stacking $\mathbf{m}_t$ along the time axis yields a SPAM $\mathbf M \in \mathbb{R}^{T\times |C|}$.

Unlike prior phonological feature maps that rely on direct target supervision \cite{king2000detection,yu2012boosting,arora2018phonological,merkx2018articulatory}, SPAM leverages the emergent representations within S3Ms \cite{choi2026b,choi2026self}.

\noindent \textbf{Silence and closure handling.}
Beyond the two channels per \texttt{PanPhon} feature, we add a dedicated silence channel and two extra channel pairs, closure ($\pm$) and release ($\pm$).
The phonological vectors described above are designed to distinguish phones, not speech and silence. 
Thus, we introduce an additional channel, \texttt{silence+}, trained by assigning silent segments to the positive set.
Also, the closure and release channels target stops and affricates, which are realized as two acoustic phenomena: a near-silent closure followed by a release burst. (TIMIT separately transcribes closure and release for stops and affricates.)
We collapse the closure and release back to the base phone during segmentation and recognition.

\subsection{Recognition head}
\label{sec:recognition-head}
Because the SPAM channels are \texttt{PanPhon} features, recognition requires no trained classifier: a segment is labeled with the phone whose phonological feature vector best matches its SPAM activations.
To do this, we precompute a \textit{canonical vector} $\mathbf{p}_v \in \mathbb{R}^{|C|}$ for each vocabulary item $v$ in the \texttt{PanPhon} phone inventory.\footnote{We restrict the inventory to segments with a defined
consonantal value ($\texttt{cons}\neq 0$), which excludes tones; we leave
tone modeling to future work.}
Each channel contains the canonical channel value for that phone, normalized by the number of active channels (each canonical vector sums to $1$). 
For each segment $s$ delimited by the segmentation head (\Cref{sec:segmentation-head}), we read the SPAM at its center frame $c(s)$ and predict the phone, $\hat{v}$, as
\begin{equation}
  \hat{v} = \arg\max_{v} \sigma({\mathbf m_{c(s)}})^{\top}\mathbf{p}_v ,
  \label{eq:recognize}
\end{equation}
where $\sigma$ applies the sigmoid function element-wise to a vector.

This amounts to a nearest-neighbor lookup in \texttt{PanPhon} and has no learned parameters.
This means that the recognizer can output phones never seen during training; extending it to a new phone requires only preparing a new canonical vector based on its phonological features, and restricting outputs to a known language's inventory requires only filtering the set of canonical vectors.
 
\subsection{Segmentation head}\label{sec:segmentation-head}
Our segmentation head is based on prominence-based peak detection~\cite{2020SciPy}, applied to an ensemble of segmentation signals derived from SPAM and a mel spectrogram.
Therefore, like the recognition head, the segmentation head is also gradient-descent-free.
We now define the individual segmentation signals:

\noindent \textbf{Difference between frames.}
Inspired by classical approaches~\cite{aversano2001new}, we place boundaries at points of maximal temporal change by measuring the cosine distance between adjacent frames of the SPAM:
\begin{equation}
  \delta_1(t) = 1 - \cos(\mathbf{m}_{t-1},\, \mathbf{m}_{t}),
  \label{eq:delta}
\end{equation}
where prominent peaks imply changes in phonological features, \textit{i.e.}, phonetic boundaries.

\noindent \textbf{Multi-scale differences.}
A single adjacent-frame distance (\Cref{eq:delta}) can be blurred by gradual transitions that unfold over several frames.
We therefore add wider differences, centered near the frame $t$:
\begin{align}
  \delta_2(t) &= 1 - \cos(\mathbf{m}_{t-1},\, \mathbf{m}_{t+1}), \\
  \delta_3(t) &= 1 - \cos(\mathbf{m}_{t-2},\, \mathbf{m}_{t+1}),
  \end{align}
which accumulate change across a broader context while keeping the peak aligned to the boundary.

\begin{table*}[tb!]
\caption{Phone segmentation performance (R-value $\uparrow$). The best and second-best scores are marked in \textbf{bold} and \underline{underlined}, respectively.
\method shows strong performance especially for unseen domains.
See \cref{sec:exp_pseg} for details.}
\label{tab:segmentation}
\vspace{-0.5em}
\centering
\resizebox{0.95\textwidth}{!}{%
\begin{tabular}{ll c c cc cc cc >{\columncolor{orange!10}}c}
\toprule
 & & \multicolumn{2}{c}{\textbf{English}} &
 \multicolumn{2}{c}{\textbf{Accented English}} &
 \multicolumn{2}{c}{\textbf{Atypical Speech}} &
 \multicolumn{2}{c}{\textbf{Multilingual}} & \textbf{Avg.} \\
\cmidrule(lr){3-4} \cmidrule(lr){5-6} \cmidrule(lr){7-8} \cmidrule(lr){9-10}
\textbf{Model} &  &
{\texttt{TIMIT}} & {\texttt{Buckeye}} &
{\texttt{GTIMIT-S}} & {\texttt{GTIMIT-T}} & 
{\texttt{TORGO}} & {\texttt{SSNCE}} &
{\texttt{Vox}} & {\texttt{GTIMIT-Thai}} & 
{\textbf{(OOD)}} \\
\midrule
\textit{Toplines} &&&&&&&&&&\\
\quad GT Transcript + MFA \cite{mcauliffe2017montreal} & & 82.4 & 84.0 & 81.2 & 83.9 & 80.1 & 53.1 & - & 66.9 & - \\
\quad KoelLabs-XLSR \cite{koellabs} + MFA \cite{mcauliffe2017montreal} & & 81.7 & 83.1 & 80.3 & 82.4 & 78.9 & - & - & - & - \\
\quad PhoneticXeus \cite{bharadwaj2026pxeus} + MFA \cite{mcauliffe2017montreal} & & 78.7 & 76.9 & 75.9 & 78.7 & 74.0 & 52.7 & - & - & - \\
\midrule
\textit{Trained on TIMIT} &&&&&&&&&&\\
\quad CTC \cite{graves2006connectionist} & & {70.6} & 71.7 & 65.7 & 62.6 & 66.4 & 52.6 & 49.2 & 65.4 & 61.9 \\
\quad FCE \cite{zhu2022charsiu} & & {\textbf{89.6}} & \textbf{80.5} & \underline{73.0} & \textbf{70.5} & 68.7 & \underline{54.3} & 63.0 & 74.1 & \underline{69.2} \\
\quad BCE \cite{strgar2023phoneme} & & {\underline{85.8}} & \underline{77.0} & 71.1 & 64.5 & \underline{71.9} & 53.4 & \underline{66.5} & \underline{75.2} & {68.5} \\
\midrule
\quad \method (Ours) & & {80.0} & 76.3 & \textbf{75.2} & \underline{68.9} & \textbf{72.2} & \textbf{60.3} & \textbf{75.1} & \textbf{75.8} & \textbf{72.0} \\
\bottomrule
\end{tabular}
}
\vspace{-1em}
\end{table*}
\noindent \textbf{Backward contrast.}
Prior work~\cite{choi2026self} showed that position-dependent phonological subspaces exist within self-supervised representations.
In other words, one S3M frame also encodes the phonological features of the \emph{preceding} phone.
Based on this, we predict the previous phone's phonological activations directly from the current S3M frame $\mathbf{r}_t$ through a least-squares regressor, which has a closed-form solution:
\begin{align}
  \bwd{\mathbf{W}} = {\arg\min}_{\textbf{W}} 
  \mathbb{E}_{l, s}
  \lVert \mathbf{W}^\top \mathbf{r}_{c(s)} - \mathbf{m}_{c(l)} \rVert^2 .
\end{align}
Here, we take all adjacent phone segment pairs $l, s$ in the training data where $l$ is a segment immediately preceding $s$.

Based on the predicted activations $\bwd{\mathbf{m}}_t = \bwd{\mathbf{W}}^\top \mathbf{r_t}$, we form a backward contrast at scale $\ell\in \{1, 2, 3\}$ and anchor offset $a = \lfloor \ell/2 \rfloor$ to keep the peak aligned to the boundary:
\begin{align}
    \beta_{\ell}(t) ={}\cos\left(\bwd{\mathbf{m}}_{t{+}a},\, \mathbf{m}_{t{+}a{-}\ell}\right) - \cos\left(\bwd{\mathbf{m}}_{t{+}a},\, \mathbf{m}_{t{+}a}\right),
\end{align}
which is large just after a phone boundary, where the predicted previous phone aligns with an earlier frame but not with the frame it is predicted from.

\noindent \textbf{Mel spectrograms.}
We also add a mel spectrogram-based signal~\cite{yang2024simple,glass1988multilevel}, independent of SPAM.
Exactly as in~\cite{yang2024simple}, we extract a log-mel spectrogram $\mathbf{F} \in \mathbb{R}^{T_\text{mel} \times 40}$ on a $10$\,ms grid and measure the cosine distance across a $4$-frame window:
\begin{align}
    \delta_\text{mel}(u) = 1 - \cos(\mathbf{F}_{u-2},\, \mathbf{F}_{u+1}),
\end{align}
where $u$ indexes the mel grid. We min-max normalize $\delta_\text{mel}$ over the utterance and subsample it onto the S3M grid by taking every other frame, so a peak at frame $t$ marks a boundary at the start of frame $t$ (consistent with other signals).

\noindent \textbf{Ensembling different segmentation signals.}
While $\delta_{1}(t)$ alone already performs well, we propose an ensemble that yields further empirical improvements (\Cref{ss:abl-seg}):
\begin{align}
    b(t)=\prod_k\left(b_k(t)-\varphi_{k}\right),
\end{align}
where $b(t)$ is the ensembled boundary signal, each $b_k$ is an individual signal, and $\varphi_k$ is its theoretical minimum, subtracted so that every signal is non-negative before the product.
The product is large where component signals agree on a change, suppressing spurious peaks.

We ensemble seven signals: multi-scale differences $\delta_1, \delta_2, \delta_3$, backward contrasts $\beta_1, \beta_2, \beta_3$, and the mel-spectrogram-based difference $\delta_\text{mel}$.
Ablations are provided in \Cref{ss:abl-seg}.
Additionally, using the silence channel, we suppress peaks falling inside silent spans.

\section{Experiments}\label{sec:expr}

We compare \method against baselines that use segmentation losses proposed in prior work, training all methods on TIMIT~\cite{garofolo1993timit} and testing out-of-domain performance on other datasets.
We also compare against state-of-the-art (SotA) systems that serve as toplines.
Our evaluations are over a wide range of datasets that reflect realistic downstream conditions, including atypical and nonnative speech and unseen languages.
We release our modeling and benchmarking code.\footnote{SPAM modeling at \url{https://github.com/juice500ml/phonespam}, evaluation at \url{https://github.com/stephenmac7/phone-metrics}, and benchmark at 
\url{https://github.com/Shikhar-S/Speech-Segmentation}.}
% \url{https://anonymous.4open.science/r/Speech-Segmentation-70AE}.

\subsection{Phone Segmentation}
\label{sec:exp_pseg}
\begin{table*}[tb!]
\caption{
Phone recognition results (PFER $\downarrow$) on the PRiSM benchmark \cite{prism}.
See \cref{sec:exp_prec} for details.
}
\vspace{-0.5em}
\label{tab:prism_pr}
\centering
\resizebox{\textwidth}{!}
{%
\begin{tabular}{ll c c cc >{\columncolor{orange!10}}c ccc >{\columncolor{orange!10}}c}
\toprule
 & & & \multicolumn{4}{c}{\textbf{Accented English Datasets}} &
 \multicolumn{4}{c}{\textbf{Multilingual Datasets}} \\
\cmidrule(lr){4-7} \cmidrule(lr){8-11}

\textbf{Model} &  & \makecell{\textbf{Learnable}\\\textbf{Param.}} & 
{\texttt{PR-tmt} \cite{garofolo1993timit} } & {\texttt{PR-arc} \cite{zhao18_interspeech} } & {\texttt{PR-saa} \cite{gmuspeechaccentarchive} }  & Avg. & {\texttt{PR-drc} \cite{paschen2020doreco} } & {\texttt{PR-vox} \cite{chodroff2024voxangeles} } & {\texttt{PR-tsm} \cite{mortensen2021tusom2021} } & Avg. \\
\midrule
\textit{Toplines} &&&&&&&&&\\
\quad KoelLabs-XLSR \cite{metzger2026scaling} & & 300M & {9.6} & {7.8} & {7.8} & {8.4} & {22.9} & {18.0} & {24.7} & {21.9} \\
\quad \pxeus \cite{bharadwaj2026pxeus}  & & 580M & 13.3 & {9.9} & {8.5} & {10.6} & {16.8} & {14.4} & {21.9} & {17.7} \\
\midrule
\textit{Trained on TIMIT} &&&&&&&&&\\
\quad CTC & & 316M & {\textbf{7.2}} & \textbf{15.3} & \textbf{11.8} & \textbf{11.4} & \textbf{20.8} & \textbf{22.4} & \textbf{22.7} & \textbf{22.0} \\
\quad FCE & & 316M & {8.7} & 18.9 & 15.8 & 14.5 & 26.4 & 40.2 & 35.9 & 34.2 \\
\midrule
\quad \method (Ours) & & 47K & {22.9} & 22.6 & 23.4 & 23.0 & 27.3 & 24.3 & 35.0 & 28.9 \\

\bottomrule
\end{tabular}
}
\vspace{-1em}
\end{table*}

\noindent\textbf{Baselines.}
We benchmark \method against the frame-wise cross-entropy (FCE)~\cite{zhu2022charsiu,xu2021explore,yeo2023speech}, frame-wise binary cross-entropy (BCE)~\cite{strgar2023phoneme}, and CTC loss~\cite{graves2006connectionist} baselines with full fine-tuning.
For a fair comparison, all use TIMIT~\cite{garofolo1993timit} as the common training data, and WavLM-large~\cite{chen2022wavlm} as the underlying S3M.

We also employ MFA~\cite{mcauliffe2017montreal} as a topline, relying on ground truth (GT) phonetic transcriptions for forced alignment.
Finally, we test cascaded systems that use state-of-the-art recognizers, like PhoneticXeus~\cite{bharadwaj2026pxeus} and KoelLabs-XLSR~\cite{metzger2026scaling, koellabs}, predicting transcripts and then force-aligning them with MFA.

\noindent\textbf{Datasets.}
We evaluate in various out-of-domain settings that cover conversational English speech (Buckeye~\cite{pitt2005buckeye}), accented English speech (GTIMIT-S~\cite{gtimit-simple} and GTIMIT-T~\cite{gtimit-tbnk}), multilingual speech (VoxAngeles~\cite{chodroff2024voxangeles}, GTIMIT-Thai~\cite{gtimit-thai}) and atypical speech (SSNCE~\cite{ssnce, ssnce-aac}, TORGO~\cite{torgo, hernandez2020dysarthria}).
Together, these cover various real-world scenarios for phone segmentation.

\noindent \textbf{Evaluation.}
We use R-value~\cite{rval} as the segmentation metric, following previous works~\cite{zhu2022charsiu,strgar2023phoneme}.
Specifically, we use a 20-ms threshold for a boundary hit along with the \texttt{strict}~\cite{strgar2023phoneme} mode of evaluation, consistent with~\cite{strgar2023phoneme}.
R-value measures how close performance is to an ideal operating point.

\noindent\textbf{Results.}
\Cref{tab:segmentation} summarizes phone segmentation performance across in-domain and out-of-domain conditions.
\method achieves the best average R-value among the baselines trained on TIMIT, indicating robust generalization.
While FCE and BCE perform competitively on English and some accented English datasets, their performance drops more substantially in atypical and multilingual settings, likely due to overfitting on TIMIT, an English dataset.
In contrast, \method performs especially well on the most challenging out-of-domain datasets, achieving the best results on GTIMIT-S, TORGO, SSNCE, VoxAngeles and GTIMIT-Thai, and remaining competitive on GTIMIT-T.

Also, comparisons against the toplines employing MFA with GT or state-of-the-art recognizers reveal that \method outperforms them on SSNCE and GTIMIT-Thai while maintaining moderate performance on English.
Note that topline systems use pretrained acoustic models (English~\cite{english_mfa_acoustic}, Tamil~\cite{Ahn_Chodroff_2022} and Thai~\cite{mfa_thai_mfa_acoustic_2024}) that are trained on far more transcribed data than \method.
Furthermore, such systems are unavailable for many languages, so cannot segment VoxAngeles, which contains 95 languages~\cite{chodroff2024voxangeles}.
The cascaded recognizers, likewise, depend on a large transcribed dataset of roughly 17{,}000 hours~\cite{zipa}.
\method, in contrast, can be applied to any language, making it more practical for low- or zero-resource languages.

\subsection{Phone Recognition} \label{sec:exp_prec}

\noindent \textbf{Setup.}
We evaluate phone recognition using the PRiSM benchmark~\cite{prism}, which covers accented English (TIMIT \cite{garofolo1993timit}, L2-ARCTIC Perceived \cite{zhao18_interspeech}, and Speech Accent Archive \cite{gmuspeechaccentarchive}) and multilingual phone recognition settings (DoReCo \cite{paschen2020doreco}, VoxAngeles \cite{chodroff2024voxangeles}, and Tusom2021 \cite{mortensen2021tusom2021}). 
Following PRiSM, we report Phone Feature Edit Rate (PFER)~\cite{powsm,mortensen2016panphon}, a soft edit-distance metric that accounts for phonological feature similarity between phones.
Unlike PRiSM, we filter out the TIMIT training split before evaluation.

\noindent\textbf{Baselines.}
To test generalization capabilities, we compare \method against FCE and CTC by training on TIMIT and evaluating on PRiSM datasets.
We also compare against SotA English and multilingual phone recognizers, KoelLabs-XLSR~\cite{metzger2026scaling, koellabs} and PhoneticXeus~\cite{bharadwaj2026pxeus}, respectively.
These models rely on high-quality human-transcribed data or large-scale G2P-based datasets for training.
In contrast, \method uses only TIMIT with pretrained speech representations, without any gradient-descent-based training.

\noindent \textbf{Results.}
\Cref{tab:prism_pr} reports phone recognition performance on PRiSM.
Among the baselines, CTC achieves the best recognition performance, followed by SPAM or FCE, even though it was the worst in segmentation.
Since CTC is optimized for recognition alone, entirely different alignments can incur the same loss.
FCE, which imposes strong supervision, degrades sharply in multilingual settings.

Interestingly, \method maintains similar performance on both accented English datasets and on TIMIT (\texttt{PR-tmt}), with only 2\% relative degradation.
CTC and FCE, by contrast, perform much better on in-domain data (TIMIT) but degrade by 60\%--117\% on accented English.
This gap is even more pronounced on multilingual datasets, where CTC and FCE degrade by 180\%--360\% while \method stays within 10\%--50\%.
These results suggest that gradient-based methods overfit to English phonotactics\footnote{The language-specific rules governing which phoneme sequences are permitted, \textit{e.g.}, English disallows words starting with [\textipa{N}].}, whereas \method is explicitly designed to avoid this, and hence generalizes more robustly to unseen languages, especially on VoxAngeles, with 95 languages across 21 language families.

Finally, the large-scale SotA toplines maintain high performance, relying on large training sets and correspondingly expensive training.
\method, in contrast, can be trained on as little as one minute of data (\Cref{ss:abl-eff}).
These results indicate that \method delivers usable performance while requiring substantially less supervision, making it more accessible to resource-scarce settings such as compute-constrained scenarios or linguistic fieldwork on endangered languages~\cite{he-etal-2024-wav2gloss}.

\begin{figure}[t]
\centering
  \includegraphics[width=0.9\linewidth]{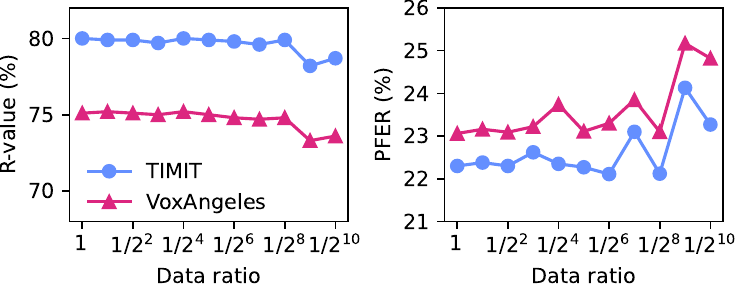}
  \vspace{-1em}
  \caption{
    Ablation on the training dataset size.
    We randomly subsample the 3-hour TIMIT training split (4{,}620 phonetically transcribed utterances) from the
    full set ($1$) down to $1/1024$.
  }
  \label{fig:efficiency}
  \vspace{-1em}
\end{figure}

\section{Analysis}

\subsection{Sample efficiency}\label{ss:abl-eff}
\noindent\textbf{Settings.} Prior work~\cite{choi2026b} showed that phonological vectors are estimated reliably even from little data, implying that \method is likewise stable.
As such, we estimate the phonological vectors from decreasing fractions of the TIMIT training set, from the full set down to $1/2, 1/4, \dots, 1/1024$.
For each fraction, we randomly subsample utterances from the 4{,}620 TIMIT training audio files.
We evaluate on TIMIT and VoxAngeles, which provide phone segmentation and are part of PRiSM~\cite{prism}.

\noindent\textbf{Results.} \Cref{fig:efficiency} shows little to no degradation as the data shrinks to $1/256$, \textit{i.e.}, only 18 utterances, with both PFER and R-value remaining on par across both datasets. Degradation is limited even at $1/1024$.
This level of sample efficiency reinforces our claim that a well-trained S3M only needs to be steered toward the task with minimal supervision, rather than fine-tuned on a large labeled dataset.
Large-scale phone recognition models such as \powsm{}, \zipactc{}, and \pxeus{} are trained on IPAPack++~\cite{zipa} with roughly 17{,}000 hours of phonetically transcribed speech, but \method can be trained with only 18 utterances ($<$ one minute) and requires negligible compute.
% Total TIMIT training data hours = 3h

\subsection{Oracle ablations}\label{ss:abl-comp}
\noindent\textbf{Ground-truth segmentation.}
To determine whether the recognizer or the segmenter is the source of the errors, we replace the predicted boundaries with ground-truth segmentation, \textit{i.e.}, a perfect segmenter, and evaluate the
recognition head alone.
Concretely, we compare the baseline SPAM (predicted segmentation) against SPAM with ground-truth (GT) segmentation.
GT segmentation lowers PFER substantially over the baseline: it more than halves the error on TIMIT (to 11.1) and cuts it to roughly a third on VoxAngeles (to 8.4).
This indicates that the recognition head is already highly capable and that the segmenter is the primary bottleneck, making it the main target for further improvement.

\noindent\textbf{Seen and unseen phones.}
We further test whether the recognition head is able to recognize phones unseen during training.
On VoxAngeles, we label each reference phone \emph{seen} if it appears in TIMIT and \emph{unseen} otherwise, and measure PFER separately on each group.
We use GT segmentation, excluding silence segments, so that each GT phone corresponds to exactly one predicted phone.
PFER is then 6.4 on seen phones and 9.4 on unseen phones.
Low PFER on unseen phones shows that SPAM recognizes unseen phones nearly as well as seen ones, likely due to the recognition head composing any phone directly from its \texttt{PanPhon} features.

\subsection{Segmentation signal ablations}\label{ss:abl-seg}
\begin{table}[t]
\centering
\caption{Phone segmentation performance comparison (R-value $\uparrow$) on different signals.}
\label{tab:seg_head}
\begin{tabular}{llcc}
\toprule
Segmentation signal && TIMIT & VoxAngeles \\
\midrule
Diff. between frames &$\delta_1$ & 79.2 & 66.1 \\
\quad + Multi-scale diff. &\quad$+ \delta_2, \delta_3$ & 77.6 & 72.7 \\
\quad + Backward contrast &\quad$+ \beta_1, \beta_2, \beta_3$ & 78.7 & 73.3 \\
\quad + Mel spectrogram &\quad$+\delta_\text{mel}$ & 80.0 & 75.1 \\
\bottomrule
\end{tabular}
  \vspace{-1em}
\end{table}

\noindent\textbf{Settings.}
We verify whether ensembling (\Cref{sec:segmentation-head}) helps and how each segmentation signal contributes.
Starting from the simplest signal $\delta_1$, we cumulatively add the multi-scale differences $\delta_2, \delta_3$, the backward contrasts $\beta_1, \beta_2, \beta_3$, and finally the mel-spectrogram difference $\delta_\text{mel}$, leading to the SPAM segmenter.

\noindent\textbf{Results.}
As shown in \Cref{tab:seg_head}, $\delta_1$ alone is already strong on TIMIT, but it is markedly weaker on VoxAngeles.
Each added signal helps a different regime: the multi-scale differences sharply improve VoxAngeles at a slight cost on TIMIT, while the backward contrasts and mel-spectrogram difference lift both datasets, yielding the best overall performance.
Thus, although the simplest single signal is already decent, ensembling signals is consistently better, especially on VoxAngeles.

% \subsection{Towards language universal phone segmentations}
% Show the bias towards English if we use TIMIT (similar to B.11 Vowel Rounding in~\cite{choi2026b}), such that we should focus on creating new datasets like VoxAngeles, which is very clean and diverse. Data cleanness is more important than scaling.

\subsection{Segmentation error analysis}
We categorize deletions (missed boundaries) by the classes of the phones they separate, and insertions (spurious boundaries) by the class of the segment they fall in. 

On TIMIT, errors are mostly deletions (Recall 73\% vs.\ Precision 86\%).
Boundaries are most often missed between vowels or between vowels and approximants (57\% missed), where we suspect activation changes are more gradual, and at silence edges (45\%).
Insertions are concentrated in stops and silences.
20\% of stops and 22\% of silences are split, compared with 4\% of vowels.
This discrepancy suggests that WavLM representations may behave differently for silence, though a detailed analysis remains for future work.

On VoxAngeles, errors are more balanced (Recall 69\% vs.\ Precision 71\%).
While per-context deletion rates are similar to TIMIT, VoxAngeles has significantly more boundaries next to silence (39\% vs.\ 9\%), leading to lower recall as \method tends to miss those.
Also, more splitting of non-silence segments accounts for over half of its lower precision.
This may stem from the longer average phone segments in VoxAngeles (132 ms vs. 83 ms).

\begin{figure}[t]
\centering
  \includegraphics[width=0.99\linewidth]{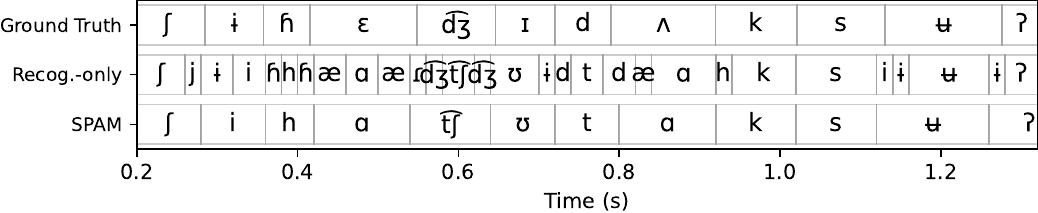}
  \vspace{-2em}
  \caption{
An example utterance from TIMIT (``She had your dark suit\ldots'').
\emph{Recog.\ only}: frame-wise predictions from the recognition head.
\emph{SPAM}: both the recognition and segmentation heads.
  }
  \label{fig:example}
  \vspace{-1em}
\end{figure}

\section{Limitations and Future Work}
\noindent\textbf{Vocabulary filter.}
The SPAM recognition head outputs narrow (fine-grained) phonetic transcriptions that reflect acoustic details (\Cref{fig:example}).
This is because the training-free recognition head is indifferent to how frequently a phone occurs in training data.
For usability, we add an optional vocabulary filter~\cite{li2020universal} that restricts the output to a known inventory of each language. We leave evaluation under known inventories or PER to future work.

\noindent\textbf{Trainable extensions.}
While we propose a gradient-descent-free approach, end-to-end fine-tuning of \method and reweighting the phonological features may improve performance.

\noindent\textbf{Feature specification.} \texttt{PanPhon} provides ternary values for each feature, which are converted into positive and negative channels.
Alternate design choices are left for future work.

\section{Conclusion}
We showed that phone segmentation and recognition can be solved jointly by reading them directly off S3M-based Phonological Activation Mapping (SPAM).
Two lightweight, gradient-descent-free heads recover phone labels and boundaries from under a minute of labeled speech, attaining strong performance on unseen phones, datasets, and languages.
This shows that well-trained S3Ms need only be steered, not retrained, for fine-grained phonetic tasks.

\section*{Acknowledgements}
This material is based upon work supported by the National Science Foundation under Grant Nos.\ 2504019 and 2238605.
This work used Delta and DeltaAI at NCSA through allocations CIS210014 and IRI120008P from the ACCESS program, supported by NSF grants \#2138259, \#2138286, \#2138307, \#2137603, and \#2138296.

\section*{AI-Generated Content Disclosure}
AI-generated content was used in preparing this manuscript, primarily for auto-completing code and revising sentences for grammar.
Nevertheless, all scientific content was conceived, verified, and finalized by the authors.

\bibliographystyle{IEEEtran}
\bibliography{refs}

\clearpage
\appendices
\section{S3M and layer ablations}\label{ss:abl-s3m}
\begin{figure}[t]
\centering
  \includegraphics[width=0.9\linewidth]{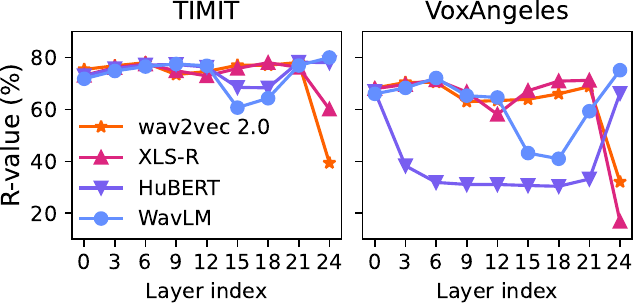}
  \caption{
    Segmentation performance on different S3Ms and layers for SPAM.
  }
  \label{fig:layerwise}
\end{figure}

\noindent\textbf{Settings.}
The 24th (final) layer of WavLM is known to be well-suited for extracting phonological vectors~\cite{choi2026b,choi2026self}, so we use it as our primary model and layer.
However, since different models and layers encode phonetic information differently~\cite{pasad2021layer,choi24b_s3mphonetic}, we sweep both.
We compare four mainstream S3Ms with the same number of transformer layers, wav2vec 2.0~\cite{baevski2020wav2vec}, XLS-R~\cite{babu22_interspeech}, HuBERT~\cite{hsu2021hubert}, and WavLM~\cite{chen2022wavlm}.

\noindent\textbf{Results.}
\Cref{fig:layerwise} shows that the final layer of WavLM achieves the best R-value on both datasets, indicating that the quality of the phonological vectors directly affects SPAM and, in turn, downstream performance.
Except for HuBERT, the layer-wise trends are broadly similar across TIMIT and VoxAngeles, though VoxAngeles amplifies the differences with sharper peaks and collapses.
Comparing XLS-R against wav2vec 2.0 shows that multilingual pretraining data does not guarantee better performance.
Note that the best model, WavLM, is an English-only pretrained model.
% HuBERT and WavLM follow similar layer-wise patterns, consistent with their close pretraining recipes, both strongest near the final layer, unlike wav2vec 2.0 and XLS-R, whose final layers degrade sharply.

\end{document}